\batchmode
\makeatletter
\def\input@path{{D:/users/ricky/Documents/academics/research/usc_madsci_papers/islg_2012/}}
\makeatother
\documentclass[oneside,english]{amsart}
\usepackage{mathpazo}
\usepackage[T1]{fontenc}
\usepackage[latin9]{inputenc}
\usepackage{array}
\usepackage{verbatim}
\usepackage{url}
\usepackage{amsthm}
\usepackage{graphicx}

\makeatletter

\providecommand{\tabularnewline}{\\}

\numberwithin{equation}{section}
\numberwithin{figure}{section}
\theoremstyle{plain}
\newtheorem{thm}{\protect\theoremname}
  \theoremstyle{definition}
  \newtheorem{defn}[thm]{\protect\definitionname}

\usepackage{graphics}
\usepackage{subfigure}

\makeatother

\usepackage{babel}
  \providecommand{\definitionname}{Definition}
\providecommand{\theoremname}{Theorem}

\begin{document}

\title{Measures of Threaded Discussion Properties}

\author{Ricky J. Sethi, Lorenzo A. Rossi, Yolanda Gil\\
USC Information Sciences Institute}
\begin{abstract}
In this paper, we present a set of measures to quantify certain properties
of threaded discussions, which are ubiquitous in online learning platforms.
In particular, we address how to measure the redundancy of posts,
the compactness of topics, and the degree of hierarchy in sub-threads.
This preliminary work would very much benefit from discussion and
serves as a starting point for ultimately creating optimal structures
of threaded discussions depending on the context.
\end{abstract}

\maketitle

\section{Introduction and Motivation}

Threaded Discussions are widely utilized in online learning platforms
like eCollege, BlackBoard, and moodle. In addition, many online forums
and Social Question \& Answer sites also rely on similar formats.
In fact, sites like \url{http://www.piazza.com}, which offer modified
threaded discussions, are being used as pedagogical supplements. Although
our ultimate goal is to be able to re-structure threaded discussions
into optimal formats depending upon the context, in this paper, we
present early stage work that seeks to quantify the characteristics
of threaded discussions as a first step.

\subsection{Overview of Related Literature }

There is a consistent literature on the automatic summarization of
textual documents by means of natural language processing (NLP) tools,
e.g. \cite{Kwon2006,Kwon2007,Ferrucci2010,Chu-Carroll2011}. Some
of the proposed approaches use automatic keyword detection to then
find out different key topics within the document. Summaries are subsequently
built by extracting the initial sentences associated with the portion
of text related to those key topics.

A subset of this literature focuses on the automatic analysis and
summarization of online single- and multi-threaded discussions.\footnote{A single-threaded discussion is an online discussion where each post
has at most one child post. In multi-threaded discussions, a post
can have more than one child post, with multiple sub-threads possible.} The focus of the application ranges from online discussions about
open source software (OSS forums) to discussions between students
attending a certain class and their instructors.

In \cite{Zhou2006}, an approach is proposed to summarize online discussions
(from the Open Source Software forum in particular). Posts are first
clustered according to topics. Then the posts belonging to each topic
are further categorized into two classes: `problem' and `advice'.
Note that, for the purpose of our research, we can look at online
single-threaded discussions as a special case of multi-threaded discussions. 

\cite{Ravi2007} study the interactions among students and teachers
in threaded discussions for distance education. The posts are classified
according to a different set subclasses called speech acts. According
to the statistics given in the paper, the majority of speech acts
turn out to be either questions or answers. The remaining ones are
elaborations of certain answers, acknowledgments, announcements, corrections
or objections. The features used for the classification are N-gram
sequences of the terms in the post (after a preliminary word filtering
stage).

Various works propose metrics to express respectively relationships
between posts and topics \cite{Lin2009}, relationship among contiguous
posts \cite{Lin2009}, coherence of the threaded discussions \cite{jihekim2006}. 

\section{Preliminary Approach }

In this document, we propose a set of measures to quantify properties
of threaded discussions (e.g. the quota of redundant posts). In the
long term, we are investigating an approach to analyze and index threaded
discussions from online learning platforms by means of machine learning
and crowd-sourcing tools. The final goal is to automatically break
down a certain threaded discussion and then be able to automatically
re-build it in ways that enhance properties valuable to a certain
target user and/or purpose: e.g. creating an automatic summary for
an instructor who needs to quickly address a students' discussion,
or a re-arrangement of the posts to the benefit of students who are
participating in the discussion. At this preliminary stage of our
research, we need to define potential optimal views of restructured
threaded discussions. This will then help us to define the desired
atomic elements in the structure of the discussions and consequently
to design the machine learning/crowd-sourcing \footnote{For instance, we could design interfaces to require contributors to
also label their own posts in a simple way before they submit it.} algorithms to break down such discussions. To help with the definition
of views, this paper proposes a set of measures to quantify some specific
properties of threaded discussions.

\subsection{Metrics }

We can represent threaded discussions with tree type data structures.
We propose the following metrics: 
\begin{itemize}
\item Redundancy of Posts
\item Topic Coherence (Compactness)
\item \textit{\emph{Degree of Hierarchy of Sub-Threads}}
\end{itemize}
We believe that these quantities can be used as simple indicators
of how \textit{good} a structure of the threaded discussion for certain
purposes is and therefore be useful for definition of desired views
of re-structured threaded discussions. Note that these measures can
be computed only over threaded discussions whose posts have been already
analyzed and classified.

Let $N$ be the number of posts $p_{i}$ in a discussion, where $i=1,\ldots,N$
and $t(p_{i})$ indicates the date and time the post was submitted. 
\begin{defn}
\textbf{Duplicate Post:} Given two posts $p_{i}$ and $p_{j}$ where
$t(p_{j})>t(p_{i})$, we say that $p_{j}$ is a duplicate post of
$p_{i}$ when the content of $p_{j}$ is so similar to the content
of $p_{i}$ that $p_{j}$ could be removed without relevant loss of
information for the readers. 

\textbf{Redundancy Factor:} Given a certain discussion, where $N_{d}$
is the total number of duplicate posts, we define the redundancy factor
as $r=N_{d}/(N-N_{d})$, where $N$ is the total number of posts in
the discussion. 
\end{defn}

In online student forums or threaded discussions, we may have participants
making posts that are duplicates of pre-existing posts in the discussion.
Under some circumstances it may be desirable to remove duplicate posts
and reduce redundancy. E.g., if we consider the original threaded
discussion in Table \ref{tab:Detailed-posts-from-Example-TD}, we
can see its redundancy with 20 total posts and 2 duplicate posts of
$r=0.11$ vs a redundancy of $r=0$ for the re-structured threaded
discussion in Table \ref{tab:Detailed-posts-from-Example-TD-restructured}.
In addition, sometimes, retaining redundancy can be useful for instructors
and users and may indicate popular topics or topic clouds. The automatic
assessment of the redundancy of a post requires a high semantic analysis
and therefore is a very challenging natural language processing task.
In our future work, we will investigate ways to infer this metric,
including using summarization methods like MEAD \footnote{\url{http://www.summarization.com/mead/}}. 

\begin{defn}
\textbf{Topic Coherence:} Let's assume that the posts of a certain
single-threaded discussion can be classified into a certain set of
topics (or stances) $s_{j}$, $j=1,\ldots,M$, where $M$ is the total
number of topics addressed in the discussion. Let $N_{j}$ the number
of posts on a topic $s_{j}$. We can map each post in the thread to
a one dimensional space, where the location in the space is simply
given by the number of parents of the post. Therefore we can measure
how dispersed (or \textbf{compact}) a certain topic is in the discussion
by measuring the standard deviation or spatial dispersion of posts
when projected to a single dimension, or possibly multiple dimensions.
\end{defn}

The automatic estimation of coherence requires a preliminary classification
of the topics addressed by the posts. A possible approach may consist
of clustering the posts based on sets of keywords. This problem will
also be a subject of our future investigations.

It may be useful in some cases to re-structure the discussion by aggregating
posts belonging to the same topic, hence the need for defining a measure
of compactness. Similarly, if we have a set of posts $p_{k}(t)$,
$k=1,...N_{j}$ belonging to a certain topic $s_{j}$, we could introduce
a measure of the distance between the chronological sequence of posts
and the sorting of posts that is most effective for user understanding
(measure of \textbf{chronological coherence}).

\begin{defn}
\textbf{Degree of Sub-Thread Hierarchy:} We can define the Degree
of Hierarchy of a sub-thread in terms of $b$ (breadth) and $d$ (depth)
as $h=d/b$. Given $M$ (number of proto-topics/stances, defined as
each first-level sub-thread) and $N_{i}$ (Number of posts in each
topic/stance, $i=1,\ldots,M$), a Flat sub-thread is when $d=1$ and
$b=N$ giving a degree of hierarchy of $h=1/N$, whereas a Hierarchical
sub-thread is when $d=max_{i}|N_{i}|$ and $b=M$, giving a degree
of hierarchy of $h=\max_{i}|N_{i}|/M$. 
\end{defn}
Thus, if we consider the original threaded discussion in Table \ref{tab:Detailed-posts-from-Example-TD},
we can see that it has a degree of hierarchy of $h=0.36$ vs a degree
of hierarchy of $h=0.80$ for the re-structured threaded discussion
in Table \ref{tab:Detailed-posts-from-Example-TD-restructured}.

\section{Conclusion}

\begin{figure}
\noindent\begin{minipage}[t][1\totalheight][b]{1\columnwidth}%
(a) \includegraphics[width=1\columnwidth]{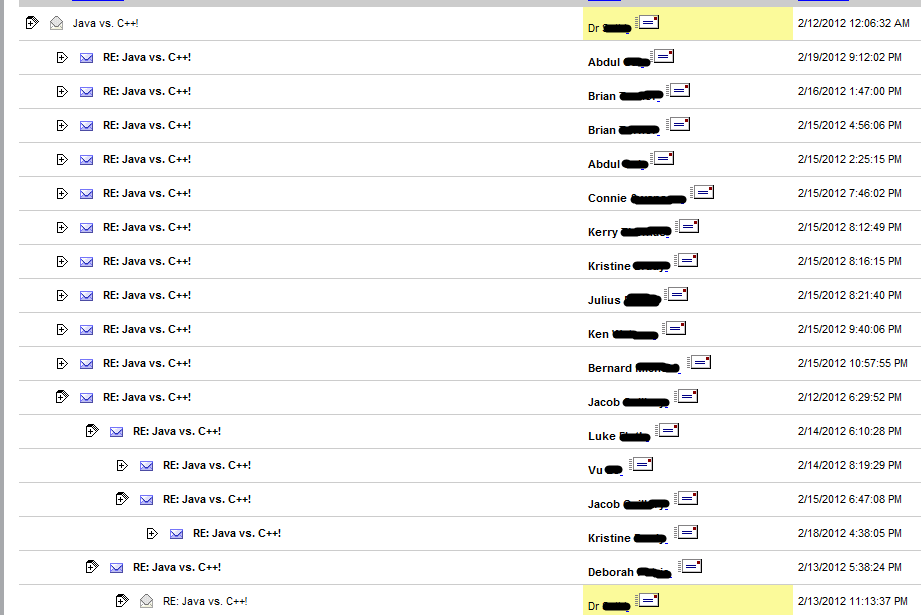}%
\end{minipage}

\noindent\begin{minipage}[t][1\totalheight][b]{1\columnwidth}%
\rule[0.5ex]{1\columnwidth}{1pt}

(b) \includegraphics[width=1\columnwidth]{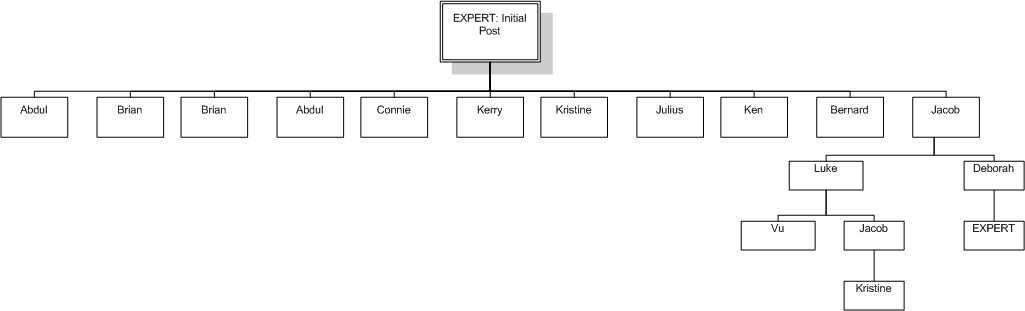}

\medskip{}
\rule[0.5ex]{1\columnwidth}{1pt}

(c) \includegraphics[width=1\columnwidth]{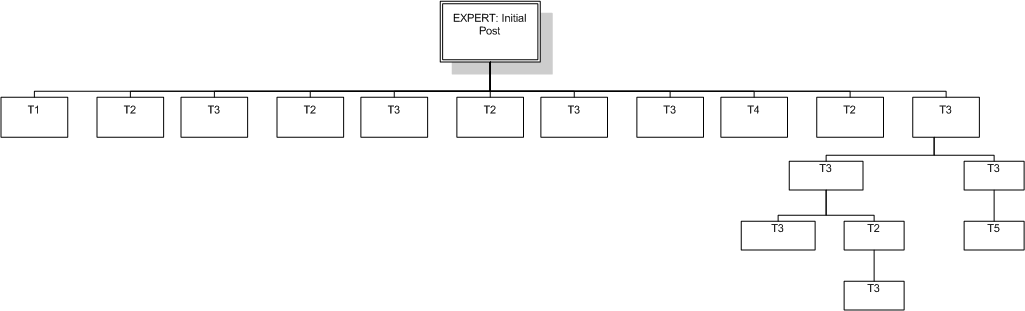}%
\end{minipage}

\caption{Example Threaded Discussion: (a) original posts, (b) tree-view of
posts, and (c) tree-view of topics. \label{fig:Example-Threaded-Discussion}}
\end{figure}

In this paper, we have presented some initial attempts to quantify
various characteristics of threaded discussions with the eventual
goal of re-structuring threaded discussions into optimal structures
depending on context. Although we have collected some threaded discussions
from online classes (examples shown in Figure \ref{fig:Example-Threaded-Discussion}),
this work is in a very early stage and we would welcome any comments
and discussion.

\bibliographystyle{plain}
\bibliography{3D__users_ricky_Documents_academics_research_us___slg_2012_references-qa_threaded_discussions}

\pagebreak{}

\appendix

\section{Appendix of Figures and Tables}

\begin{table}
\begin{tabular}{>{\raggedright}p{0.7\columnwidth}>{\raggedright}p{0.3\columnwidth}}
\hline 
\textbf{Post}  &
\textbf{Topic} \tabularnewline
\hline 
{\scriptsize{}Abdul: To the Bonus question: C++ design aspect is very
limited...} &
{\scriptsize{}T1 }\tabularnewline
\hline 
{\scriptsize{}Brian: The syntax of Java is closely based on the syntax
of the C++ programming language... } &
{\scriptsize{}T2 }\tabularnewline
\hline 
{\scriptsize{}Brian: Although there are many similarities I believe
using the Java language is a bit easier.} &
{\scriptsize{}T3}\tabularnewline
\hline 
{\scriptsize{}Abdul: There are so many differences and similarities,
it sometimes is hard to mistake...} &
{\scriptsize{}T2}\tabularnewline
\hline 
{\scriptsize{}Connie: When Sun Microsystems came out with Java, it
like an answer to MOST of their prayers...} &
{\scriptsize{}T3}\tabularnewline
\hline 
{\scriptsize{}Kerry: I prefer the one that's right for the job. For
a high-level user-facing application...} &
{\scriptsize{}T2}\tabularnewline
\hline 
{\scriptsize{}Kristine: three things that come to mind on what is
different between Java \& C++ are...} &
{\scriptsize{}T3}\tabularnewline
\hline 
{\scriptsize{}Julius: I do find JAVA to be a bit less combersum when
putting together your methods...} &
{\scriptsize{}T3}\tabularnewline
\hline 
{\scriptsize{}Ken: Why might it be said that Java is an object-oriented
language while C++ is a procedural...} &
{\scriptsize{}T4}\tabularnewline
\hline 
{\scriptsize{}Bernard: I think that both have there place.} &
{\scriptsize{}T2}\tabularnewline
\hline 
\begin{itemize}
\item {\scriptsize{}Jacob: ... With Java, the programming is more user friendly...}

\begin{itemize}
\item {\scriptsize{}Luke: I love the analogy, Jacob...}

\begin{itemize}
\item {\scriptsize{}Vu: Hi Luke\textendash I had...}{\scriptsize \par}
\item {\scriptsize{}Jacob: ... I took the same classes...}

\begin{itemize}
\item {\scriptsize{}Kristine: Java has both kinds...}{\scriptsize \par}
\end{itemize}
\end{itemize}
\item {\scriptsize{}Deborah: I don't think Java...}

\begin{itemize}
\item {\scriptsize{}EXPERT: Sounds like Java has 1 and C++...}

\begin{itemize}
\item {\scriptsize{}Deborah: The STL is the Standard Template...}{\scriptsize \par}
\item {\scriptsize{}Ajay: What is STL? What is C++'s STL...}{\scriptsize \par}
\end{itemize}
\item {\scriptsize{}Jody: I agree Deborah, and I may be because I...}
\end{itemize}
\end{itemize}
\end{itemize}
 &
\begin{itemize}
\item {\scriptsize{}T3}

\begin{itemize}
\item {\scriptsize{}T3}

\begin{itemize}
\item {\scriptsize{}T3}{\scriptsize \par}
\item {\scriptsize{}T2}

\begin{itemize}
\item {\scriptsize{}T3}{\scriptsize \par}
\end{itemize}
\end{itemize}
\item {\scriptsize{}T3}

\begin{itemize}
\item {\scriptsize{}T5}

\begin{itemize}
\item {\scriptsize{}T5}{\scriptsize \par}
\item {\scriptsize{}T5}{\scriptsize \par}
\end{itemize}
\item {\scriptsize{}T3}
\end{itemize}
\end{itemize}
\end{itemize}
\tabularnewline
\hline 
 &
\tabularnewline
\end{tabular}

\caption{Detailed posts from Example, un-structured Threaded Discussion in
Figure \ref{fig:Example-Threaded-Discussion}. Here, we see the hierarchical
posts on the left and the corresponding Topic for each post on the
right, such that all posts belong to one of five topics, T1 - T5.
\label{tab:Detailed-posts-from-Example-TD} }
\end{table}

\begin{table}
\begin{tabular}{>{\raggedright}p{0.7\columnwidth}>{\raggedright}p{0.3\columnwidth}}
\hline 
\textbf{Post}  &
\textbf{Topic} \tabularnewline
\hline 
\begin{itemize}
\item {\scriptsize{}Abdul: To the Bonus question: C++ design aspect is very
limited...}
\end{itemize}
 &
\begin{itemize}
\item {\scriptsize{}T1 }
\end{itemize}
\tabularnewline
\hline 
\begin{itemize}
\item {\scriptsize{}Brian: The syntax of Java is closely based on the syntax
of the C++ programming language... }

\begin{itemize}
\item {\scriptsize{}Abdul: There are so many differences and similarities,
it sometimes is hard to mistake... }{\scriptsize \par}
\item {\scriptsize{}Kerry: I prefer the one that's right for the job. For
a high-level user-facing application... }{\scriptsize \par}
\item {\scriptsize{}Bernard: I think that both have there place. }

\begin{itemize}
\item {\scriptsize{}Jacob: ... I took the same classes...}
\end{itemize}
\end{itemize}
\end{itemize}
 &
\begin{itemize}
\item {\scriptsize{}T2 }

\begin{itemize}
\item {\scriptsize{}T2 }{\scriptsize \par}
\item {\scriptsize{}T2 }{\scriptsize \par}
\item {\scriptsize{}T2 }

\begin{itemize}
\item {\scriptsize{}T2}
\end{itemize}
\end{itemize}
\end{itemize}
\tabularnewline
\hline 
\begin{itemize}
\item {\scriptsize{}Julius: I do find JAVA to be a bit less combersum when
putting together your methods... }

\begin{itemize}
\item {\scriptsize{}Connie: When Sun Microsystems came out with Java, it
like an answer to MOST of their prayers... }{\scriptsize \par}
\item {\scriptsize{}Kristine: three things that come to mind on what is
different between Java \& C++ are... }{\scriptsize \par}
\item {\scriptsize{}Jacob: ... With Java, the programming is more user friendly...
}

\begin{itemize}
\item {\scriptsize{}Luke: I love the analogy, Jacob... }

\begin{itemize}
\item {\scriptsize{}Vu: Hi Luke\textendash I had... }{\scriptsize \par}
\item {\scriptsize{}Kristine: Java has both kinds... }{\scriptsize \par}
\end{itemize}
\item {\scriptsize{}Deborah: I don't think Java... }

\begin{itemize}
\item {\scriptsize{}Jody: I agree Deborah, and I may be because I...}
\end{itemize}
\end{itemize}
\end{itemize}
\end{itemize}
 &
\begin{itemize}
\item {\scriptsize{}T3 }

\begin{itemize}
\item {\scriptsize{}T3 }{\scriptsize \par}
\item {\scriptsize{}T3 }{\scriptsize \par}
\item {\scriptsize{}T3 }

\begin{itemize}
\item {\scriptsize{}T3 }

\begin{itemize}
\item {\scriptsize{}T3 }{\scriptsize \par}
\item {\scriptsize{}T3 }{\scriptsize \par}
\end{itemize}
\item {\scriptsize{}T3 }

\begin{itemize}
\item {\scriptsize{}T3}
\end{itemize}
\end{itemize}
\end{itemize}
\end{itemize}
\tabularnewline
\hline 
\begin{itemize}
\item {\scriptsize{}Ken: Why might it be said that Java is an object-oriented
language while C++ is a procedural...}
\end{itemize}
 &
\begin{itemize}
\item {\scriptsize{}T4}
\end{itemize}
\tabularnewline
\hline 
\begin{itemize}
\item {\scriptsize{}EXPERT: Sounds like Java has 1 and C++...}

\begin{itemize}
\item {\scriptsize{}Deborah: The STL is the Standard Template...}
\end{itemize}
\end{itemize}
 &
\begin{itemize}
\item {\scriptsize{}T5}

\begin{itemize}
\item {\scriptsize{}T5 }
\end{itemize}
\end{itemize}
\tabularnewline
\hline 
 &
\tabularnewline
\end{tabular}

\caption{Restructured representation of the Threaded Disucssion from Table
\ref{tab:Detailed-posts-from-Example-TD}, after removing redundant
posts and minimizing the dispersion of topics. Here, T1-T5 again represent
five different topics (stances) inferred from the posts. Different
levels of minimization of these properties would be desirable under
different conditions. E.g., this can be useful to students who might
have a hard time distinguishing important topics or components when
they're dispersed or serve as a reference which summarizes the content
instead of having dispersed content that makes it difficult to understand
the central ideas. \label{tab:Detailed-posts-from-Example-TD-restructured} }
\end{table}

\end{document}